\documentclass[seceq,supplement]{ptptex} 
\usepackage{graphicx} 
 
\title{Strangeness nuclear physics -- 2010} 
\subtitle{Overview of strangeness nuclear physics} 
\author{Avraham \textsc{Gal}\footnote{ e-mail address: 
avragal@vms.huji.ac.il}} 
\inst{Racah Institute of Physics, The Hebrew University, 
Jerusalem 91904, Israel}

\abst{Selected topics in Strangeness Nuclear Physics are reviewed: 
$\Lambda$-hypernuclear spectroscopy and structure, 
multistrangeness, and $\overline K$ mesons in nuclei.} 

\begin{document} 

\maketitle

\section{Introduction}
\label{sec:intro}

The properties of hypernuclei reflect the nature of the underlying
baryon-baryon interactions and, thus, can provide tests of models 
for the free-space hyperon-nucleon ($YN$) and hyperon-hyperon ($YY$) 
interactions. The Nijmegen group has constructed a number of meson-exchange, 
soft-core models, using SU(3)$_{\rm f}$ symmetry to relate coupling constants 
and form factors \cite{rijken10}. The J\"ulich group, in addition to $YN$ 
meson exchange models \cite{haidenbauer05}, published recently leading-order 
chiral effective-field theory $YN$ and $YY$ potentials \cite{polinder06}. 
Quark models have also been used within the $(3q)-(3q)$ resonating group 
model (RGM), augmented by a few effective meson exchange potentials of scalar 
and pseudoscalar meson nonets directly coupled to quarks\cite{fujiwara07}. 
Finally, we mention recent lattice QCD calculations \cite{beane07,nemura09}. 

On the experimental side, there is a fair amount of data on single-$\Lambda$ 
hypernuclei, including production, structure and decay modes \cite{hashim06}. 
Little is known on strangeness $S$=$-2$ hypernuclei. The missing information 
is vital for extrapolating into strange hadronic matter \cite{sbg93} (SHM) 
for both finite systems and in bulk, and into neutron stars \cite{jsb08}. 
Therefore, following a review of the spectroscopy of single-$\Lambda$ 
hypernuclei in Sect.~\ref{sec:lambda}, I update in Sect.~\ref{sec:LL} what is 
known about $\Lambda\Lambda$ hypernuclei and discuss in Sect.~\ref{sec:shm} 
the nuclear potential depths anticipated for other hyperons ($\Sigma,\Xi$) 
from $YN$ interaction models, as well as from the scarce hypernuclear data 
available for these hyperons. Aspects of $\overline K$ nuclear interactions 
are reviewed in Sect.~\ref{sec:Kbar}, highlighting the issue of kaon 
condensation.

\section{$\Lambda$ hypernuclei} 
\label{sec:lambda} 

\begin{figure}[tbh] 
\centering
\includegraphics[width=6.0cm]{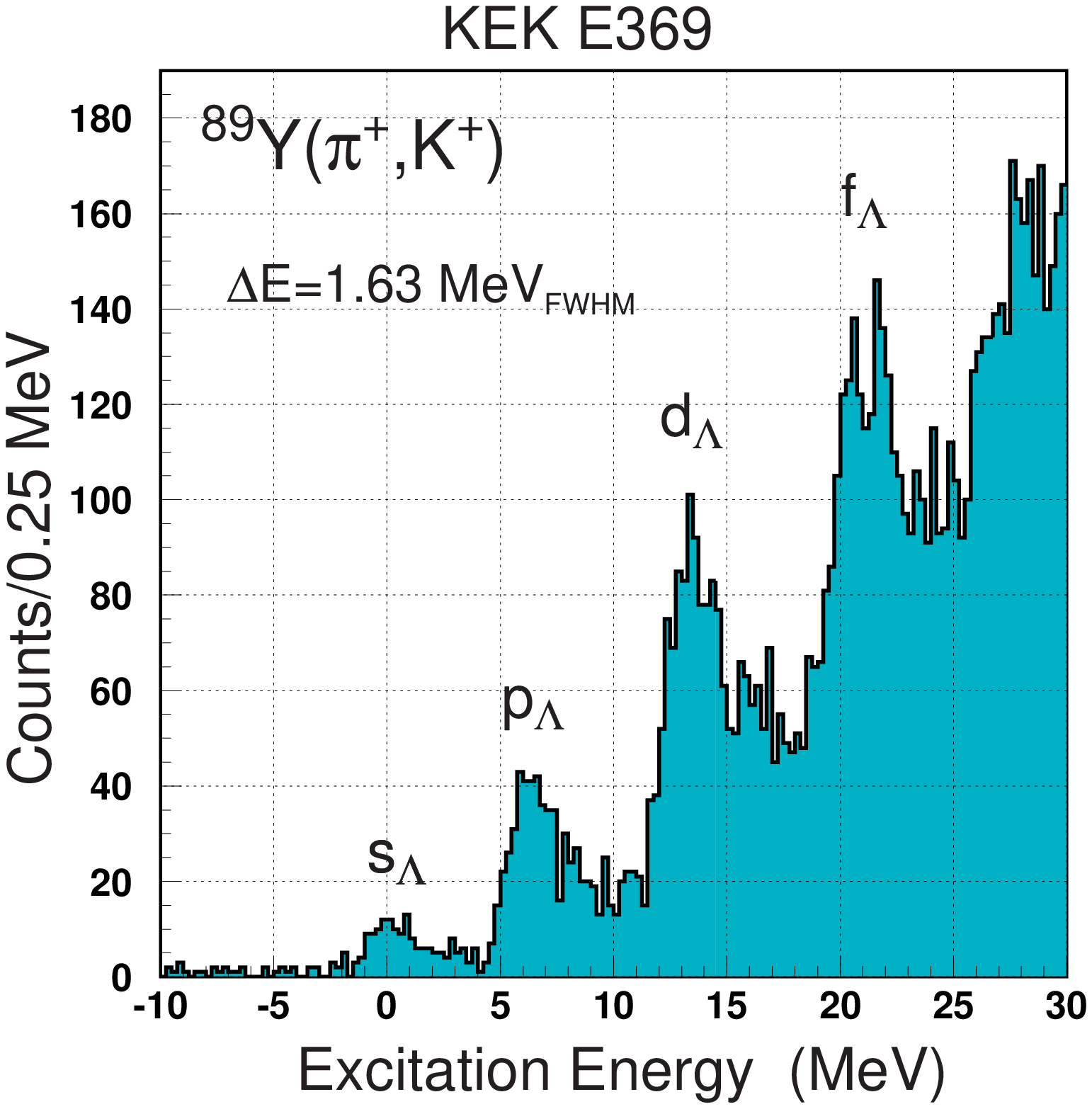} 
\hspace*{3mm} 
\includegraphics[width=7.5cm]{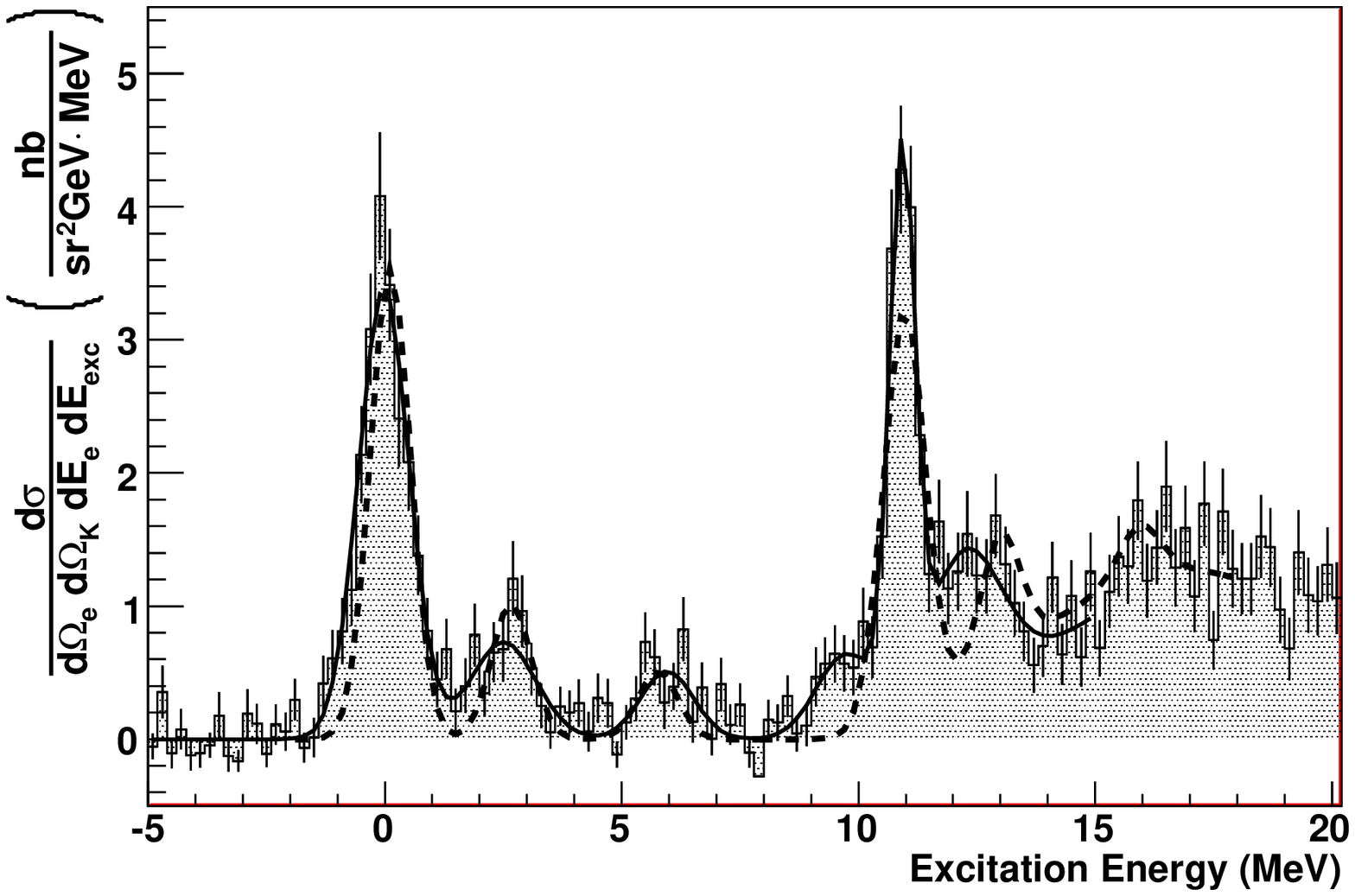} 
\caption{Left: $(\pi^+,K^+)$ spectrum of $_{~\Lambda}^{89}{\rm Y}$ 
from KEK E369~\cite{hotchi01}. Right: ($e,e'K^+$) spectrum of 
$_{~\Lambda}^{12}{\rm B}$ from Jlab~\cite{cusanno10}.} 
\label{fig:s.p.}
\end{figure} 

To test $YN$ models against the considerable body of information on $\Lambda$ 
hypernuclei, effective interactions for use in limited spaces of shell-model 
orbits must be evaluated. The $\Lambda$ well depth resulting from soft-core 
Nijmegen nuclear-matter $G$-matrices \cite{rijken10} can be brought to 
a reasonable agreement with the empirical value 28 MeV deduced in fitting 
binding energies of $\Lambda$ single-particle (s.p.) states \cite{mdg88}. 
However, the partial-wave contributions, in particular the spin dependence 
of the central interaction, vary widely in different models, and the 
$\Lambda$-nuclear spin-orbit splittings do not come sufficiently small in 
a natural way in most of the available models.{\footnote{Nevertheless, it was 
suggested recently \cite{finelli09} that $\Lambda\to\Sigma\to\Lambda$ iterated 
one-pion exchange contributions overlooked in MF approaches cancel out the 
short-range $\sigma + \omega$ MF contributions to the $\Lambda$ nuclear 
spin-orbit potential.}} The l.h.s. of Fig.~\ref{fig:s.p.} shows one of the 
most impressive examples of $\Lambda$ s.p. structure.  Although the structure 
of the $f_{\Lambda}$ orbit in $_{~\Lambda}^{89}{\rm Y}$ may suggest 
a spin-orbit splitting of 1.7 MeV, a more careful shell-model analysis 
demonstrates consistency with a $\Lambda$ spin-orbit splitting of merely 
0.2 MeV, with most of the observed splitting due to mixing of 
$\Lambda N^{-1}$ particle-hole excitations \cite{motoba08}. 
Interesting hypernuclear structure is also revealed between major $\Lambda$ 
s.p. states in $_{~\Lambda}^{12}{\rm C}$. This has not been studied yet with 
sufficient resolution in medium-weight and heavy hypernuclei, but data already 
exist from JLab on $^{12}$C and other targets, with sub-MeV resolution, 
as shown on the r.h.s. of Fig.~\ref{fig:s.p.}. Furthermore, even with the 
coarser resolution of the $(\pi^+,K^+)$ data shown in Fig.~\ref{fig:s.p.}, 
most of the $_{~\Lambda}^{12}{\rm C}$ levels between the (left) $1s_{\Lambda}$ 
peak and the (right) $1p_{\Lambda}$ peak are particle-stable and could be 
studied by looking for their electromagnetic cascade deexcitation to the 
ground state. 

A systematic program of $\gamma$-ray measurements \cite{tamura08} has been 
carried out for light $\Lambda$ hypernuclei at BNL and KEK in order to study 
the spin dependence of the {\it effective} $\Lambda N$ interaction in the 
nuclear $p$ shell \cite{dg78},  
\begin{equation} 
V_{\Lambda N} = \bar{V}+\Delta\,{\vec s}_N\cdot {\vec s}_\Lambda+S_\Lambda\, 
{\vec l}_N\cdot {\vec s}_\Lambda + S_N\,{\vec l}_N\cdot {\vec s}_N + T\,S_{12} 
\,, 
\label{eq:spin} 
\end{equation} 
specified here by four radial matrix elements: $\Delta$ for spin-spin, 
$S_\Lambda$ and $S_N$ for spin-orbit, $T$ for the tensor interaction. 
The most completely studied hypernucleus todate is $_{\Lambda}^7{\rm Li}$ with 
five observed $\gamma$-ray transitions, allowing a good determination of these 
parameters in the beginning of the $p$ shell \cite{millener08}: 
\begin{equation} 
A=7,9: \,\,\,\,\,\, 
\Delta=430,\,\, S_\Lambda=-15,\,\, S_N=-390,\,\, T=30 \,\,\,({\rm keV}). 
\label{eq:7-9} 
\end{equation} 
The dominant spin-dependent contributions to $_{\Lambda}^7{\rm Li}$ are 
due to $\Delta$ for $1s_\Lambda$ inter-doublet spacings listed in 
Table~\ref{tab:doublet}, and to $S_N$ for intra-doublet spacings. 

\begin{figure}[tbh] 
\centering 
\includegraphics[width=13cm]{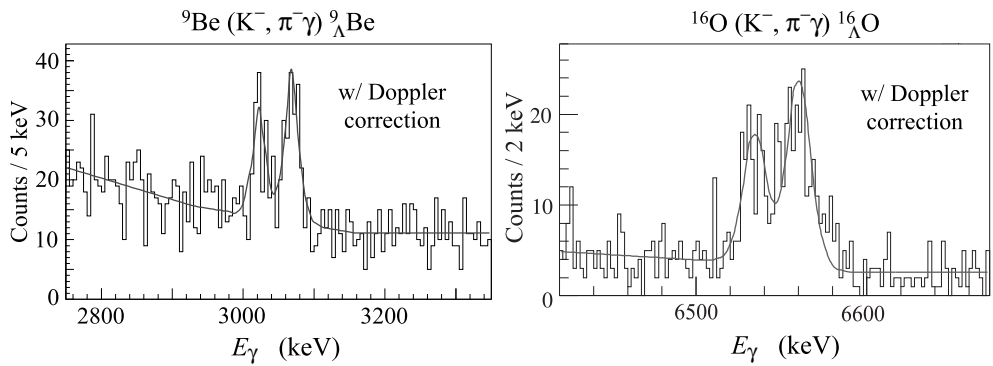}
\caption{$\gamma$-ray spectra of $\Lambda$ hypernuclei from BNL E930, see 
Tamura's review~\cite{tamura08}. The observed twin peaks (in order left to 
right) result from the ${\frac{5}{2}}^+$ and ${\frac{3}{2}}^+$ levels in 
$_{\Lambda}^9{\rm Be}$ separated by 43 keV, deexciting to the ground state, 
and from deexcitation of a $1^{-\star}$ level in $_{~\Lambda}^{16}{\rm O}$ 
to the ground-state doublet $0^-$ and $1^-$ levels separated by 26 keV. } 
\label{fig:9be16o}
\end{figure}

A remarkable experimental observation of minute doublet spin splittings 
in $_{\Lambda}^9{\rm Be}$ and in $_{~\Lambda}^{16}{\rm O}$ is shown in 
Fig.~\ref{fig:9be16o}. The contributions of the various spin-dependent 
components in Eq.~(\ref{eq:spin}) to these and other doublet splittings 
are given in Table~\ref{tab:doublet} using Eq.~(\ref{eq:7-9}) for 
$_{\Lambda}^9{\rm Be}$ and a somewhat revised parameter set for heavier 
hypernuclei which exhibit greater sensitivity, in the $p_{\frac{1}{2}}$ 
subshell, to the tensor interaction \cite{millener08}: 
\begin{equation} 
A>9: \,\,\,\,\,\,
\Delta=330,\,\, S_\Lambda=-15,\,\, S_N=-350,\,\, T=23.9 \,\,\,({\rm keV}). 
\label{eq:11-15} 
\end{equation} 
Listed also are $\Lambda\Sigma$ mixing contributions, from 
Ref.~\citen{millener08}. Core polarization contributions normally bounded by 
10 keV are not listed. In $_{\Lambda}^9{\rm Be}$, since both $\Delta$ and 
$T$ are well controlled by data from other systems, it is fair to state that 
the observed $43\pm 5$ keV doublet splitting provides a stringent measure of 
the smallness of $S_{\Lambda}$, the $\Lambda$ spin-orbit parameter for 
$1s_{\Lambda}$ states, consistently with the small $\Lambda$ spin-orbit 
splitting $152\pm 54({\rm stat.})\pm 36({\rm syst.})$ keV observed in 
$_{~\Lambda}^{13}{\rm C}$ between the $1p_{\frac{1}{2}} \to {\rm g.s.}$ and 
$1p_{\frac{3}{2}} \to {\rm g.s.}$ $\gamma$-ray transitions \cite{ajimura01}. 
It is worth noting that some of the hypernuclear energy shifts observed, 
with respect to core states in the middle of the nuclear $1p$ shell, are 
poorly understood, requiring perhaps $\Lambda NN$ contribution beyond the 
substantial shifts occasionally provided by the induced nuclear spin-orbit 
parameter $S_N$ \cite{millener08}. 

\begin{table}
\caption{Calculated $1s_{\Lambda}$ doublet splitting 
contributions~\cite{millener08} $E_j$, $j$ running over $\Lambda\Sigma$ mixing 
and $\Lambda N$ spin-dependent interaction terms, using Eq.~(\ref{eq:7-9}) for 
$_{\Lambda}^7{\rm Li}$ and $_{\Lambda}^9{\rm Be}$, and Eq.~(\ref{eq:11-15}) 
for $A>9$. The calculated total splittings $E_{\rm calc}$ are compared with 
$E_{\rm exp}$ from experiment~\cite{tamura08} (in keV).} 
\label{tab:doublet} 
\begin{center} 
\begin{tabular}{lccccccccc}\hline\hline 
$_{\Lambda}^{Z}A$ & $J_{\rm upper}$ & $J_{\rm lower}$ & $E_{\Lambda\Sigma}$ & 
$E_{\Delta}$ & $E_{S_\Lambda}$ & $E_{S_{\rm N}}$ & $E_{T}$ & $E_{\rm calc}$ &
$E_{\rm exp}$ \\ \hline  
$_{\Lambda}^7{\rm Li}$ & ${\frac{3}{2}}^+$ & ${\frac{1}{2}}^+$ & $72$ & $628$ 
& $-1$ & $-4$ & $-9$ & $693$ & $692$ \\ 
$_{\Lambda}^7{\rm Li}$ & ${\frac{7}{2}}^+$ & ${\frac{5}{2}}^+$ & $74$ & $557$ 
& $-32$ & $-8$ & $-71$ & $494$ & $471$ \\ 
$_{\Lambda}^9{\rm Be}$ & ${\frac{3}{2}}^+$ & ${\frac{5}{2}}^+$ & $-8$ & $-14$ 
& $37$ & $0$ & $28$ & $44$ & $43\pm 5$ \\ 
$_{~\Lambda}^{11}{\rm B}$ & ${\frac{7}{2}}^+$ & ${\frac{5}{2}}^+$ & $56$ & 
$339$ & $-37$ & $-10$ & $-80$ & $267$ & $263$ \\ 
$_{~\Lambda}^{11}{\rm B}$ & ${\frac{3}{2}}^+$ & ${\frac{1}{2}}^+$ & $61$ & 
$424$ & $-3$ & $-44$ & $-10$ & $475$ & $504$ \\ 
$_{~\Lambda}^{12}{\rm C}$ & $2^-$ & $1^-$ & $61$ & $175$ & $-12$ & $-13$ & 
$-42$ & $153$ & $161$ \\ 
$_{~\Lambda}^{15}{\rm N}$ & ${\frac{1}{2}}^+$ & ${\frac{3}{2}}^+$ & $42$ & 
$232$ & $34$ & $-8$ & $-208$ & $92$ & \\ 
$_{~\Lambda}^{15}{\rm N}$ & ${\frac{3}{2}}_2^+$ & ${\frac{1}{2}}_2^+$ & 
$65$ & $451$ & $-2$ & $-16$ & $-10$ & $507$ & $481$ \\  
$_{~\Lambda}^{16}{\rm O}$ & $1^-$ & $0^-$ & $-33$ & $-123$ & $-20$ & $1$ & 
$188$ & $23$ & $26.4 \pm 1.7$ \\ 
$_{~\Lambda}^{16}{\rm O}$ & $2^-$ & $1_2^-$ & $92$ & $207$ & $-21$ & $1$ & 
$-41$ & $248$ & $224$ \\ \hline 
\end{tabular}
\end{center} 
\end{table} 

\begin{figure}[tbh]
\centering  
\includegraphics[width=6.7cm]{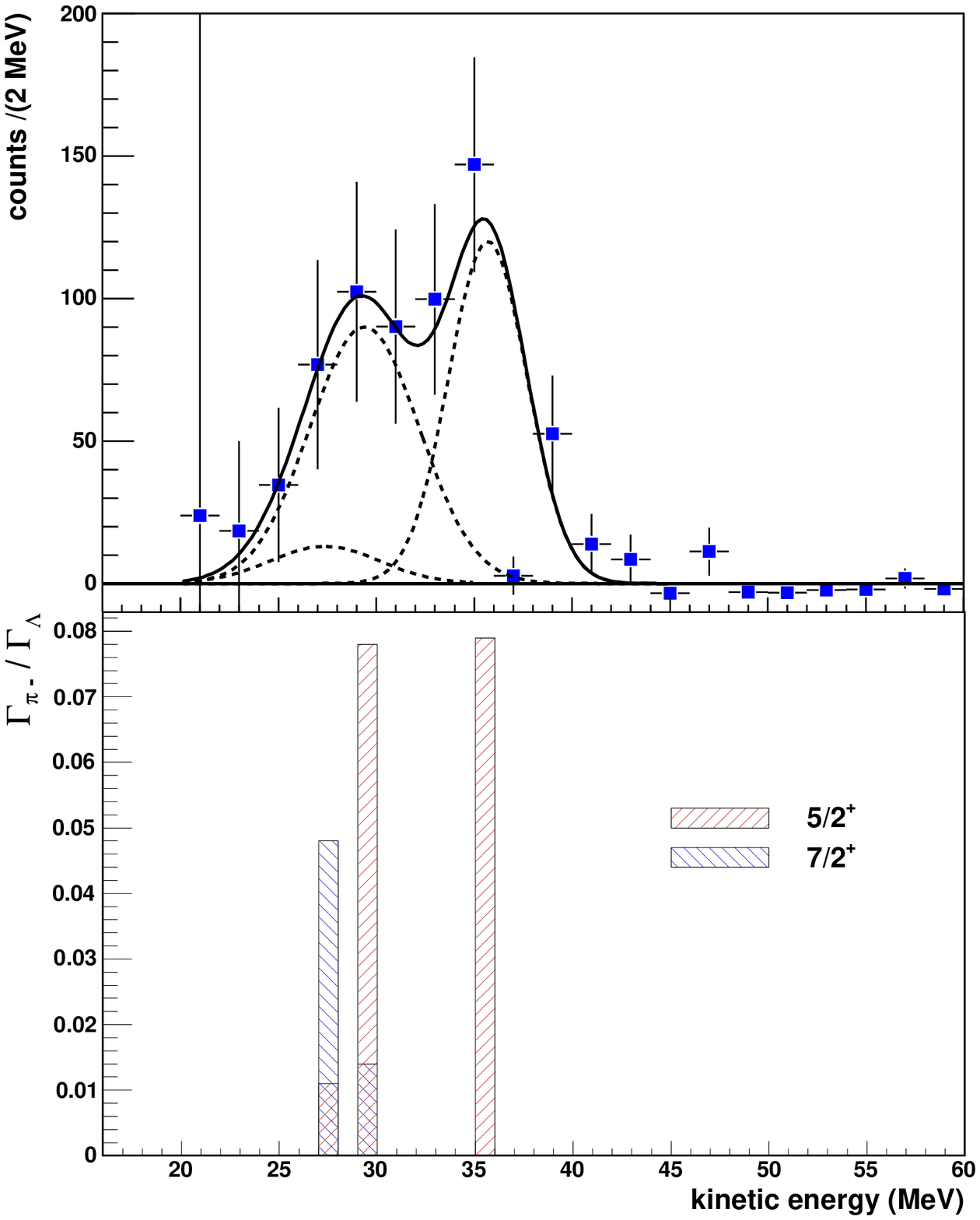} 
\hspace*{3mm} 
\includegraphics[width=6.7cm]{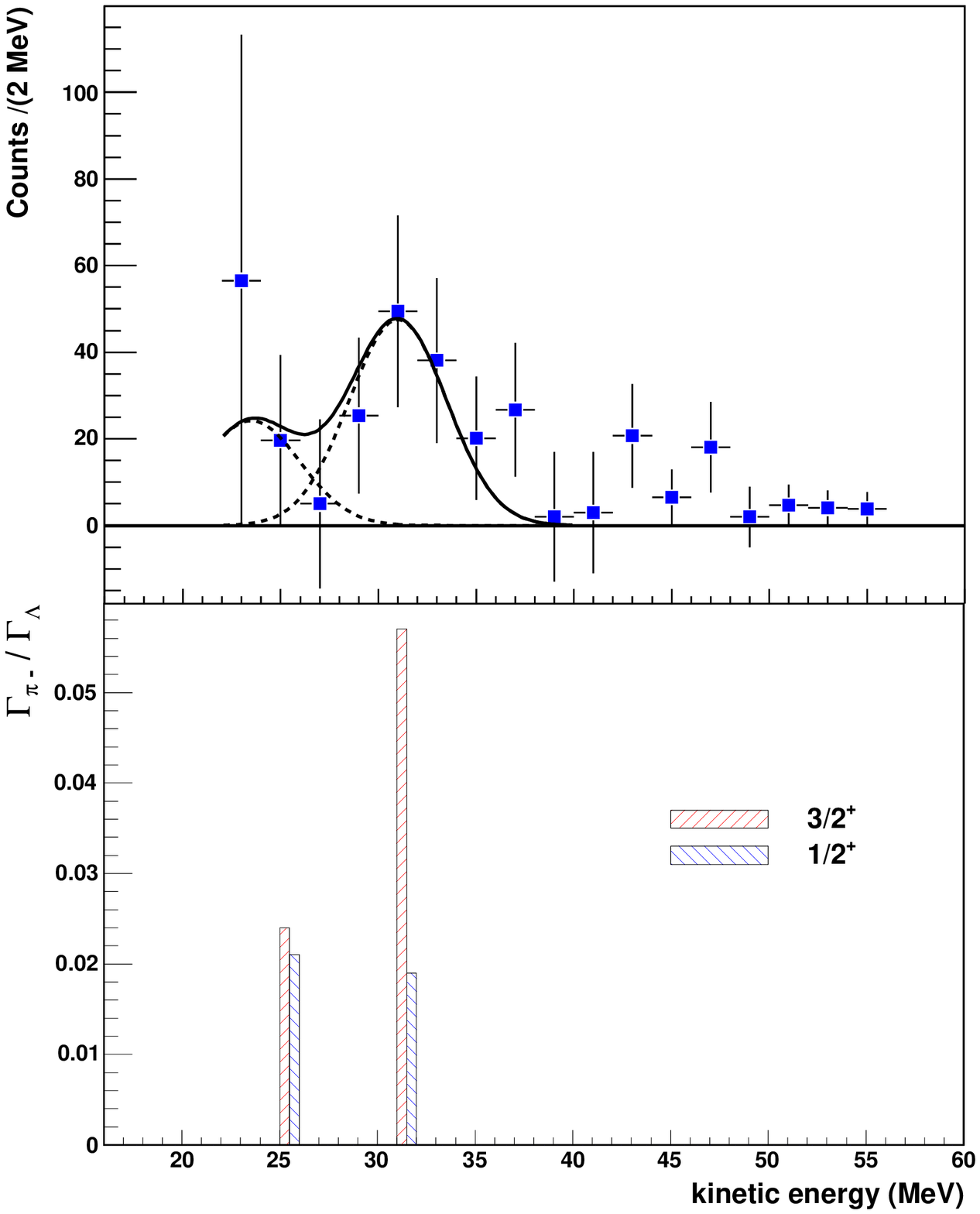}
\caption{Left: $_{~\Lambda}^{11}{\rm B} \to \pi^- + {^{11}{\rm C}}$ 
weak decay spectrum, with spin-parity $5/2^+$ favored over $7/2^+$. 
Right: $_{~\Lambda}^{15}{\rm N} \to \pi^- + {^{15}{\rm O}}$ weak decay 
spectrum, with spin-parity $3/2^+$ favored 
over spin-parity $1/2^+$. Spectra taken by FINUDA~\cite{finuda09}.} 
\label{fig:spin}
\end{figure}

The spin dependence of the $\Lambda N$ interaction may also be studied, 
as reported recently by the FINUDA Collaboration \cite{finuda09}, observing 
pionic weak decay spectra. This is demonstrated in Fig.~\ref{fig:spin} 
for two species. The $_{~\Lambda}^{11}{\rm B} \to \pi^- + {^{11}{\rm C}}$ 
spectrum shown on the l.h.s. confirms the spin-parity assignment already 
established for $_{~\Lambda}^{11}{\rm B}_{\rm g.s.}$, 
$J^{\pi}(_{~\Lambda}^{11}{\rm B}_{\rm g.s.})={\frac{5}{2}}^+$, whereas  
the $_{~\Lambda}^{15}{\rm N}\to\pi^-+{^{15}{\rm O}}$ spectrum shown on the 
r.h.s. suggests a spin-parity assignment 
$J^{\pi}(_{~\Lambda}^{15}{\rm N}_{\rm g.s.})={\frac{3}{2}}^+$, consistently 
with the positive value listed in Table~\ref{tab:doublet} for the ground-state 
doublet splitting \cite{millener08} $E({\frac{1}{2}}^+)-E({\frac{3}{2}}^+)$. 

\section{$\Lambda\Lambda$ hypernuclei} 
\label{sec:LL} 

\begin{table} 
\caption{Compiled binding energies (in MeV) of $\Lambda\Lambda$ 
hypernuclei \cite{nakazawa10}, and as calculated \cite{hiyama10,galmil10} 
fitting $V_{\Lambda\Lambda}$ to 
$B_{\Lambda\Lambda}(_{\Lambda\Lambda}^{~~6}{\rm He})_{\rm exp}$. 
The Hida event is assigned either to $_{\Lambda\Lambda}^{~11}{\rm Be}$ or to 
$_{\Lambda\Lambda}^{~12}{\rm Be}$. Values of $\delta B_{\Lambda\Lambda}
(_{\Lambda\Lambda}^{~~\rm A}{\rm Z}) \equiv B_{\Lambda\Lambda}
(_{\Lambda\Lambda}^{~~\rm A}{\rm Z})_{\rm calc} - B_{\Lambda\Lambda}
(_{\Lambda\Lambda}^{~~\rm A}{\rm Z};~V_{\Lambda\Lambda}=0)_{\rm calc}$ 
(in MeV) are given in brackets for the calculation of Ref. \citen{hiyama10}.} 
\label{tab:LL} 
\begin{center} 
\begin{tabular}{ccccc} \hline\hline 
event & $_{\Lambda\Lambda}^{~~\rm A}{\rm Z}$ & 
$B_{\Lambda\Lambda}(_{\Lambda\Lambda}^{~~\rm A}{\rm Z})_{\rm exp}$ & 
$B_{\Lambda\Lambda}(_{\Lambda\Lambda}^{~~\rm A}{\rm Z})_{\rm calc}$ 
\cite{hiyama10} & $B_{\Lambda\Lambda}
(_{\Lambda\Lambda}^{~~\rm A}{\rm Z})_{\rm calc}$ \cite{galmil10}  \\ \hline 
E373-Nagara\cite{nagara01} & $_{\Lambda\Lambda}^{~~6}{\rm He}$ & 
$6.91\pm 0.16$ & $6.91~(0.54)$ & $6.91\pm 0.16$ \\ 
Danysz {\it et al.}\cite{danysz63} & $_{\Lambda\Lambda}^{~10}{\rm Be}$ 
& $14.8\pm 0.4$ & $14.74~(0.53)$ & $15.08\pm 0.20$ \\ 
E373-Hida\cite{nakazawa10} & $_{\Lambda\Lambda}^{~11}{\rm Be}$ & $20.83\pm 
1.27$ & $18.23~(0.56)$ & $18.39\pm 0.27$  \\ 
E373-Hida\cite{nakazawa10} & $_{\Lambda\Lambda}^{~12}{\rm Be}$ & 
$22.48\pm 1.21$ & -- & $20.71\pm 0.20$ \\ 
E176\cite{aoki91} & $_{\Lambda\Lambda}^{~13}{\rm B}$ & $23.3\pm 0.7$ & -- & 
$23.21\pm 0.21$ \\ 
\hline 
\end{tabular}  
\end{center}  
\end{table} 

\begin{figure}[tbh] 
\centering 
\includegraphics[width=6.5cm,height=7.0cm]{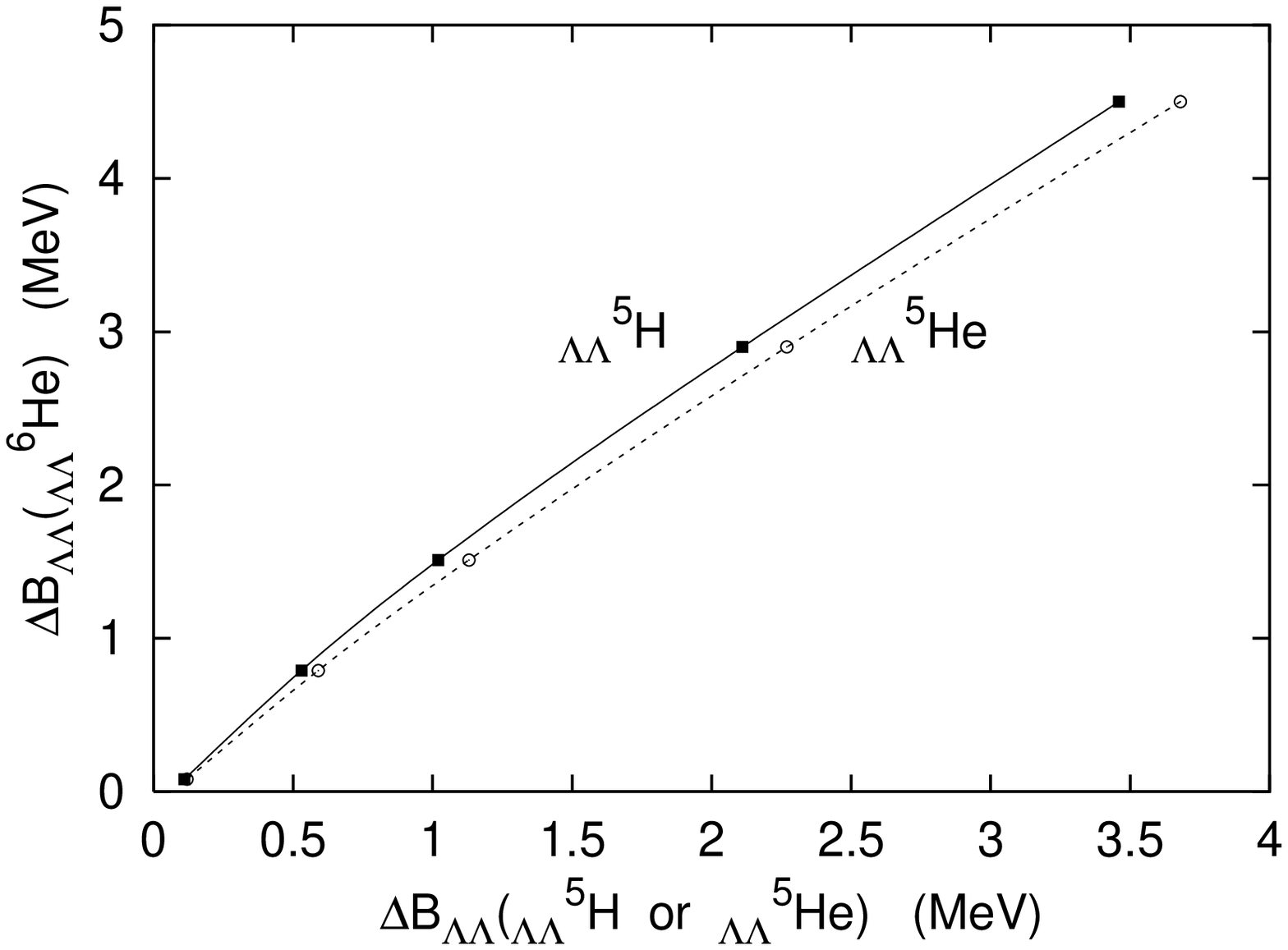} 
\hspace*{3mm} 
\includegraphics[width=6.5cm,height=7.0cm]{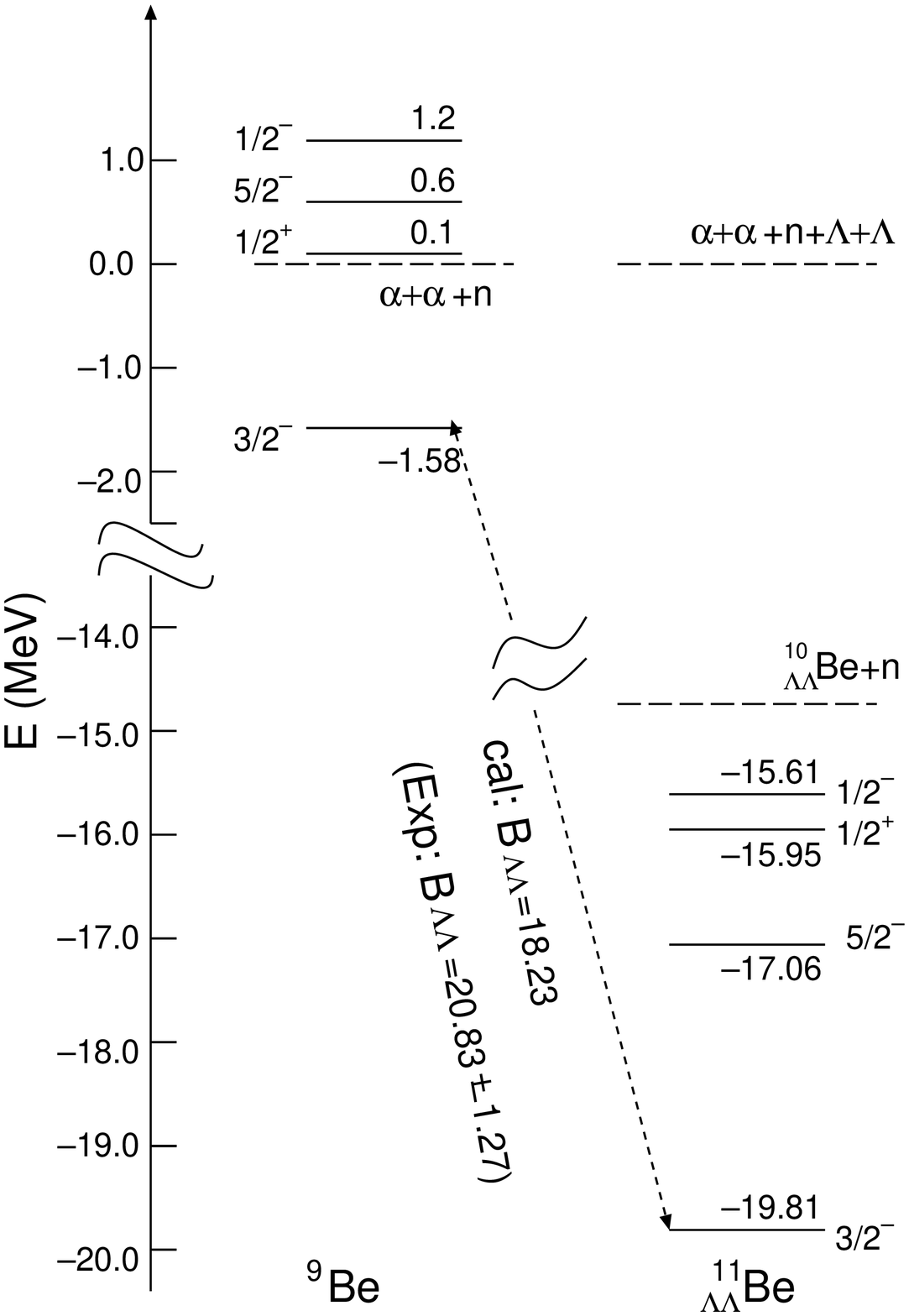} 
\caption{Left: FY relationship between 
$\Delta B_{\Lambda\Lambda}(_{\Lambda\Lambda}^{~~6}{\rm He})$ and 
$\Delta B_{\Lambda\Lambda}(_{\Lambda\Lambda}^{~~5}{\rm H}$ or 
$_{\Lambda\Lambda}^{~~5}{\rm He})$ (Filikhin and Gal \cite{fg02}). 
Right: spectrum of $_{\Lambda\Lambda}^{~11}{\rm Be}$ in a 5-cluster 
calculation \cite{hiyama10}.} 
\label{fig:LL} 
\end{figure}

Several $\Lambda\Lambda$ hypernuclear assignments have been proposed 
based on $\Xi^-$ capture events observed in hybrid emulsion KEK 
experiments \cite{nakazawa10}, as listed in Table~\ref{tab:LL}. In particular, 
the Nagara event \cite{nagara01} yields a unique assignment and a ground-state 
binding energy value for $_{\Lambda\Lambda}^{~~6}{\rm He}$ which is the 
lightest particle stable $\Lambda\Lambda$ hypernucleus established so far; 
a claim for $_{\Lambda\Lambda}^{~~4}{\rm H}$ from BNL-E906 \cite{ahn01} has 
been downgraded recently \cite{hunger07}. Comprehensive stochastic variational 
calculations of $s$-shell $\Lambda$ and $\Lambda\Lambda$ hypernuclei by Nemura 
{\it et al.} \cite{nemura05}, using meson-exchange based phenomenological 
coupled-channel potentials to account for $\Lambda N-\Sigma N$ and 
$\Lambda\Lambda-\Xi N$ mixings, predict that $_{\Lambda\Lambda}^{~~4}{\rm H}$ 
and $_{\Lambda\Lambda}^{~~5}{\rm H}$ - $_{\Lambda\Lambda}^{~~5}{\rm He}$ are 
also particle stable. These predictions depend quantitatively on the assumed 
value of $\Delta B_{\Lambda\Lambda}(_{\Lambda\Lambda}^{~~6}{\rm He}) \equiv 
B_{\Lambda\Lambda}(_{\Lambda\Lambda}^{~~6}{\rm He}) - 
2B_{\Lambda}(_{\Lambda}^{5}{\rm He})$ which now stands on 
$0.67\pm 0.17$~MeV \cite{nakazawa10} rather than slightly over $1$~MeV 
in these calculations, and since $_{\Lambda\Lambda}^{~~4}{\rm H}$ is 
calculated to be particle stable only by 2 keV, it is likely to be 
particle unstable in agreement also with a Faddeev-Yakubovsky (FY) 
calculation \cite{filikhin02}. In contrast, the particle stability 
of $_{\Lambda\Lambda}^{~~5}{\rm H}$ and $_{\Lambda\Lambda}^{~~5}{\rm He}$, 
which have not yet been discovered, appears theoretically robust \cite{fg02}, 
as shown on the l.h.s. of Fig.~\ref{fig:LL} as a function of the strength 
assumed for $V_{\Lambda\Lambda}$. We therefore conclude that the onset of 
$\Lambda\Lambda$ hypernuclear binding is likely to occur for $A=5$. 

Hiyama {\it et al.} \cite{hiyama02,hiyama10} reported on cluster calculations 
of $\Lambda\Lambda$ hypernuclei beyond $_{\Lambda\Lambda}^{~~6}{\rm He}$. 
The good agreement between 
$B_{\Lambda\Lambda}(_{\Lambda\Lambda}^{~10}{\rm Be})_{\rm exp}$ and 
$B_{\Lambda\Lambda}(_{\Lambda\Lambda}^{~10}{\rm Be})_{\rm calc}$ in 
Table~\ref{tab:LL} rests on the assumption that 
$_{\Lambda\Lambda}^{~10}{\rm Be}_{\rm g.s.}$ was identified \cite{danysz63} 
by its $\pi^-$ decay to $_{\Lambda}^{9}{\rm Be}^{\ast}(3~{\rm MeV})$. 
It is consistent with the production of 
$_{\Lambda\Lambda}^{~10}{\rm Be}^{\ast}_{2^+}(\sim 3~{\rm MeV})$ 
in the Demachi-Yanagi event \cite{demachiyanagi01}. For 
$_{\Lambda\Lambda}^{~11}{\rm Be}$, a calculated spectrum is shown on the 
r.h.s. of Fig.~\ref{fig:LL}. The $\approx 2\sigma$ discrepancy between 
$B_{\Lambda\Lambda}(_{\Lambda\Lambda}^{~11}{\rm Be})_{\rm exp}$ 
and $B_{\Lambda\Lambda}(_{\Lambda\Lambda}^{~11}{\rm Be})_{\rm calc}$ 
casts doubts on this interpretation of the Hida event. For heavier species, 
for which only shell-model simple estimates are available \cite{galmil10}, it 
is seen that the $_{\Lambda\Lambda}^{~12}{\rm Be}$ interpretation of Hida is 
as dubious as $_{\Lambda\Lambda}^{~11}{\rm Be}$. 
The good agreement between 
$B_{\Lambda\Lambda}(_{\Lambda\Lambda}^{~13}{\rm B})_{\rm exp}$ and 
$B_{\Lambda\Lambda}(_{\Lambda\Lambda}^{~13}{\rm B})_{\rm calc}$ in 
Table~\ref{tab:LL} rests on the assumption that 
$_{\Lambda\Lambda}^{~13}{\rm B}_{\rm g.s.}$ was identified \cite{aoki91} 
by its $\pi^-$ decay to $_{~\Lambda}^{13}{\rm C}^{\ast}(4.9~{\rm MeV})$.

\section{$\Lambda,\Sigma,\Xi$ hyperon nuclear potential depths and SHM} 
\label{sec:shm} 

A vast body of $(K^-,\pi^{\pm})$ and $(\pi^-,K^+)$ data indicate 
a repulsive $\Sigma$ nuclear potential, with a substantial isospin 
dependence \cite{dmg89} which for very light nuclei may conspire in 
selected configurations to produce $\Sigma$ hypernuclear quasibound states. 
The most recent $(K^-,\pi^{\pm})$ spectra \cite{bart99}, plus the very recent 
$(\pi^-,K^+)$ spectra \cite{noumi02} and related DWIA analyses \cite{kohno04}, 
suggest that $\Sigma$ hyperons do not bind in heavier nuclei. 

\begin{figure}[tbh] 
\centering 
\includegraphics[width=6.5cm]{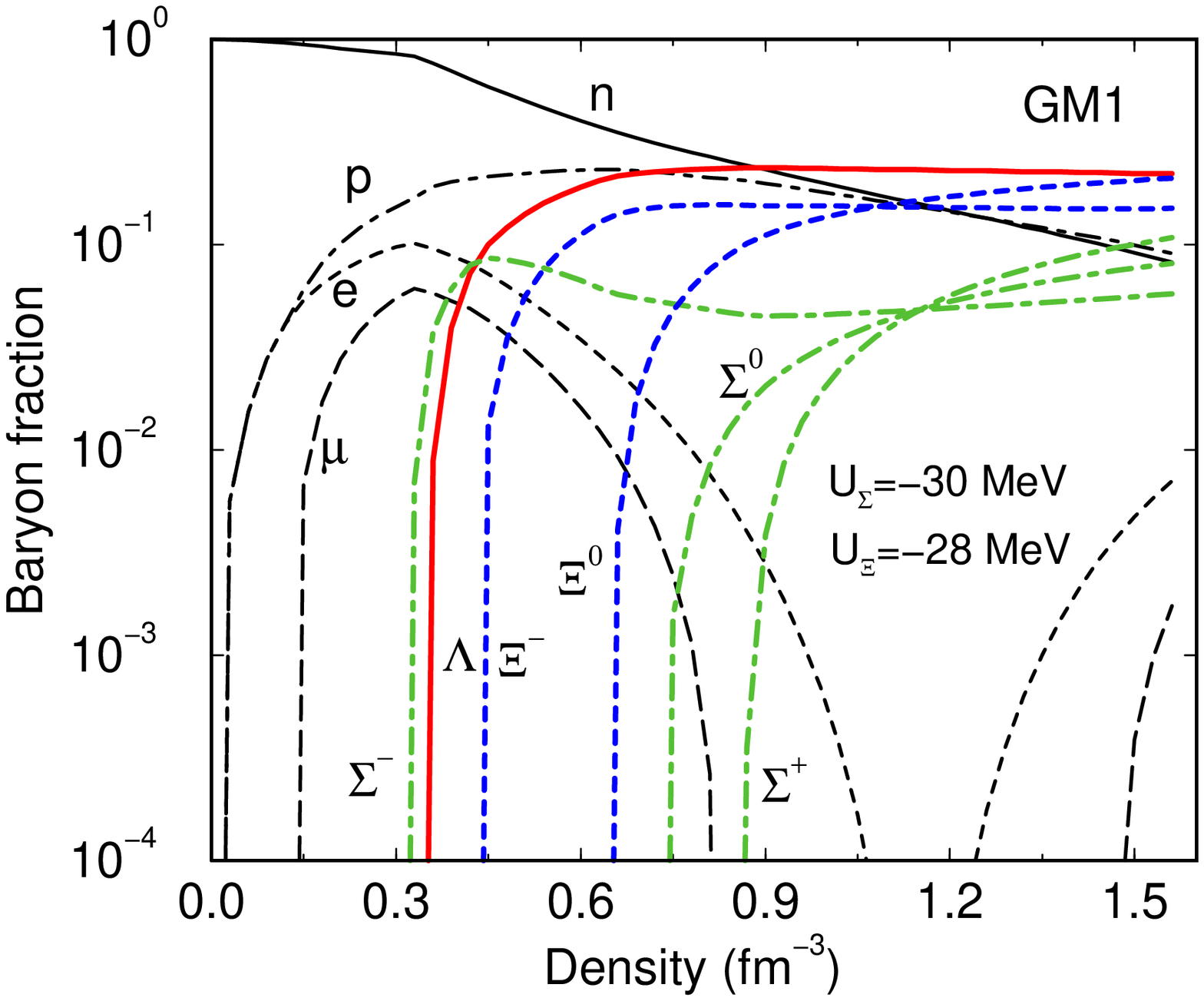} 
\hspace*{3mm} 
\includegraphics[width=6.5cm]{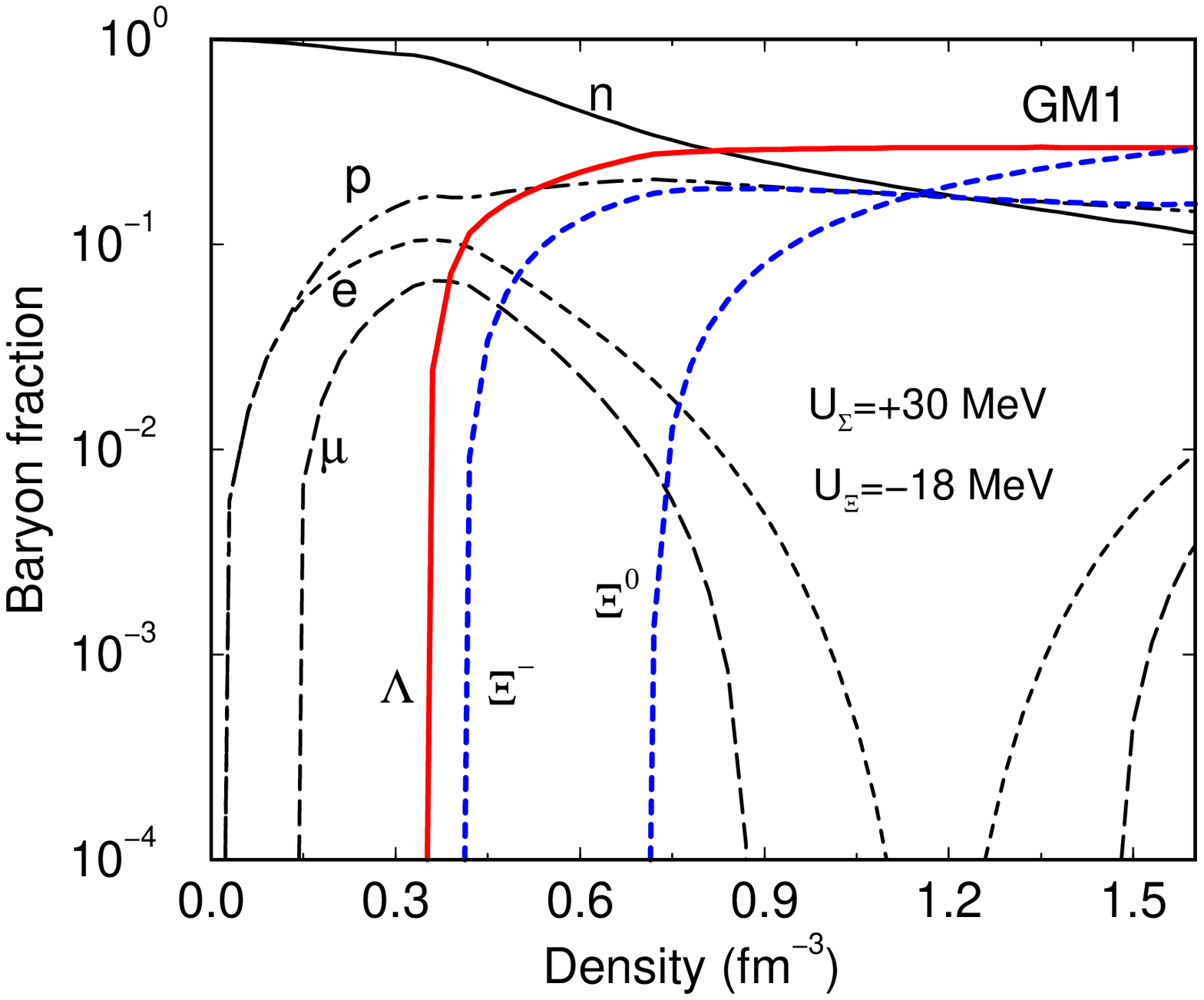} 
\caption{Left: fractions of baryons and leptons in neutron star matter 
calculated for two scenarios of hyperon nuclear potentials in RMF with 
weak $YY$ potentials \cite{jsb08}.} 
\label{fig:sbg} 
\end{figure}

A repulsive component of the $\Sigma$ nuclear potential arises also from 
analyzing strong-interaction shifts and widths in $\Sigma^-$ atoms \cite{fg07}. 
In fact, Re~$V_{\rm opt}^{\Sigma}$ is attractive at low densities 
outside the nucleus, as enforced by the observed `attractive' $\Sigma^-$ 
atomic level shifts, changing into repulsion on approach of the nuclear 
radius. The precise size of the repulsive component within the nucleus, 
however, is model dependent \cite{bfg94,mfgj95}. This repulsion bears 
interesting consequences for the balance of strangeness in the inner crust 
of neutron stars, primarily by delaying to higher densities, or even aborting 
the appearance of $\Sigma^-$ hyperons, as shown in Fig.~\ref{fig:sbg}. 

\begin{table} 
\caption{Isoscalar and isovector hyperon potentials, Eq.~(\ref{eq:lane}) 
in MeV, calculated for Nijmegen soft-core potential models 
\cite{rijken10}, denoted by year and version, at $k_F=1.35~{\rm fm}^{-1}$ 
corresponding to nuclear-matter density. Excluded are Im~$V^{\Sigma}$ due 
to $\Sigma N\to \Lambda N$ and Im~$V^{\Xi}$ due to $\Xi N\to \Lambda\Lambda$.} 
\label{tab:sig} 
\begin{center} 
\begin{tabular}{lccccccc} \hline\hline 
& 97f & 04d & 06d & 08a & 08b & phenom. & Ref. \\ \hline
$V_0^{\Lambda}$ & $-31.7$ & $-44.1$ & $-44.5$ & $-35.6$ & $-34.0$ & 
$-28$ & \citen{mdg88} \\ 
$V_0^{\Sigma}$ & $-13.9$ & $-26.0$ & $-1.2$ & $+13.4$ & $+20.3$ & 
$+30\pm 20$ & \citen{mfgj95,kohno04} \\ 
$V_1^{\Sigma}$ & $-30.4$ & $+30.4$ & $+52.6$ & $+64.5$ & $+85.2$ & 
$\approx +80$ & \citen{dgm84} \\ 
$V_0^{\Xi}$ & & $-18.7$ & & $-20.2$ & $-32.4$ & 
$\approx -14$ & \citen{fukuda98} \\ 
$V_1^{\Xi}$ & & $+50.9$ & & $-40.4$ & $-69.7$ & & \\ \hline 
\end{tabular} 
\end{center} 
\end{table} 

The $G$-matrices constructed from Nijmegen soft-core potential models 
have progressed throughout the years to produce $\Sigma$ repulsion in 
symmetric nuclear matter, as demonstrated in Table~\ref{tab:sig} using 
the parametrization 
\begin{equation} 
\label{eq:lane} 
V^Y = V_0^Y + \frac{1}{A}~V_1^Y~{\bf T}_A{\cdot}{\bf t}_Y ~. 
\end{equation} 
In the latest Nijmegen ESC08 model \cite{rijken10}, this repulsion is 
dominated by repulsion in the isospin $T=3/2,~{^3S_1}-{^3D_1}~\Sigma N$ 
coupled channels where a strong Pauli exclusion effect is suggested by SU(6) 
quark-model RGM calculations \cite{fujiwara07}. A strong repulsion appears 
also in a recent SU(3) chiral perturbation calculation \cite{kaiser05} which 
yields $V_0^{\Sigma}\approx 60$~MeV. Phenomenologically $V_0^{\Sigma} > 0$ 
and $V_1^{\Sigma} > 0$, as listed in the table, and both components of 
$V^{\Sigma}$ give repulsion in nuclei.{\footnote{In the case of $^4_\Sigma$He, 
the only known quasibound $\Sigma$ hypernucleus \cite{hayano89,nagae98}, the 
isovector term provides substantial attraction owing to the small value of $A$ 
towards binding the $T=1/2$ hypernuclear configuration, while the isoscalar 
repulsion reduces the quasibound level width \cite{harada98}.}} 

\begin{figure}[tbh] 
\centering 
\includegraphics[width=6.5cm]{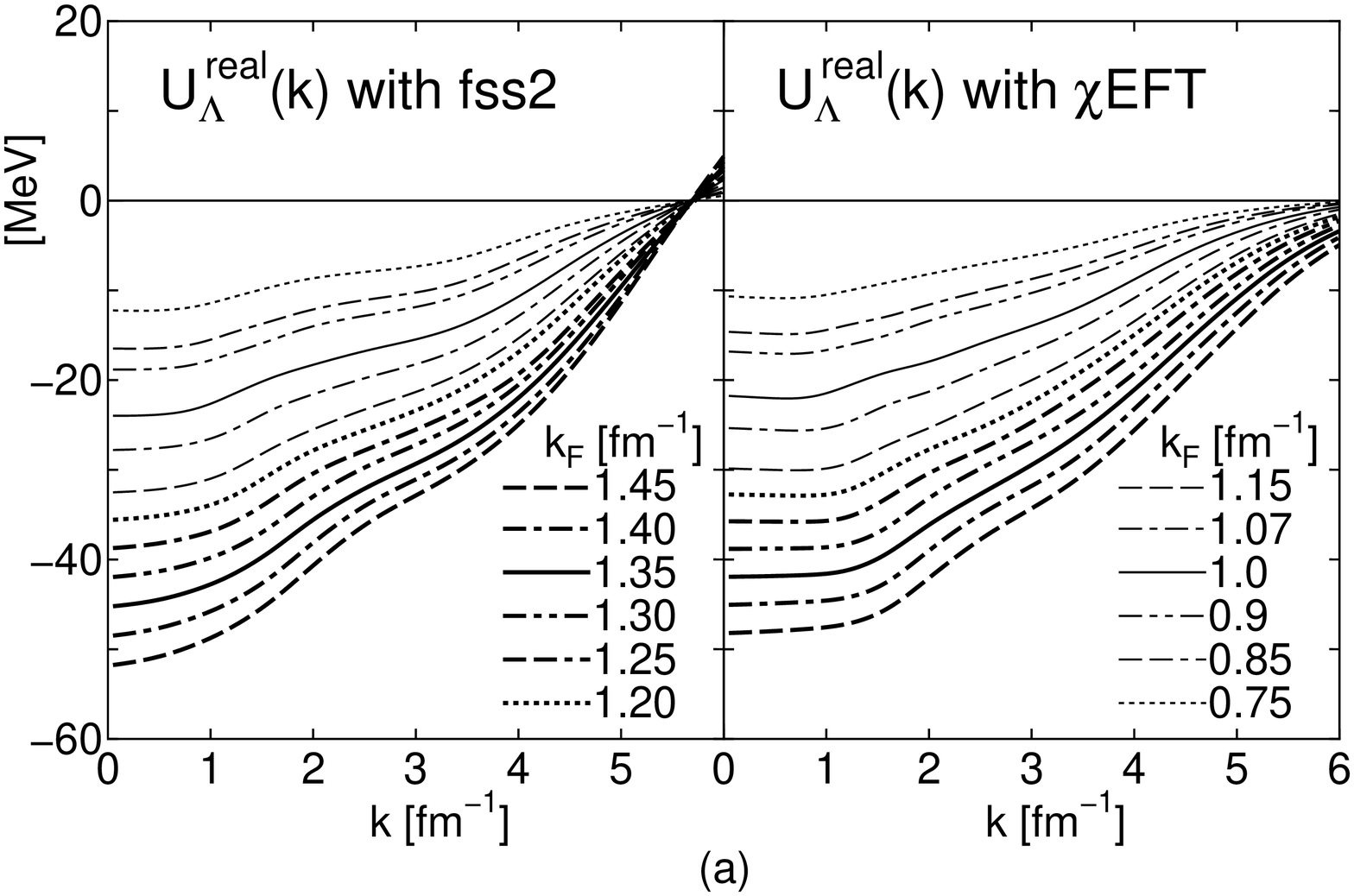} 
\hspace*{3mm} 
\includegraphics[width=6.5cm]{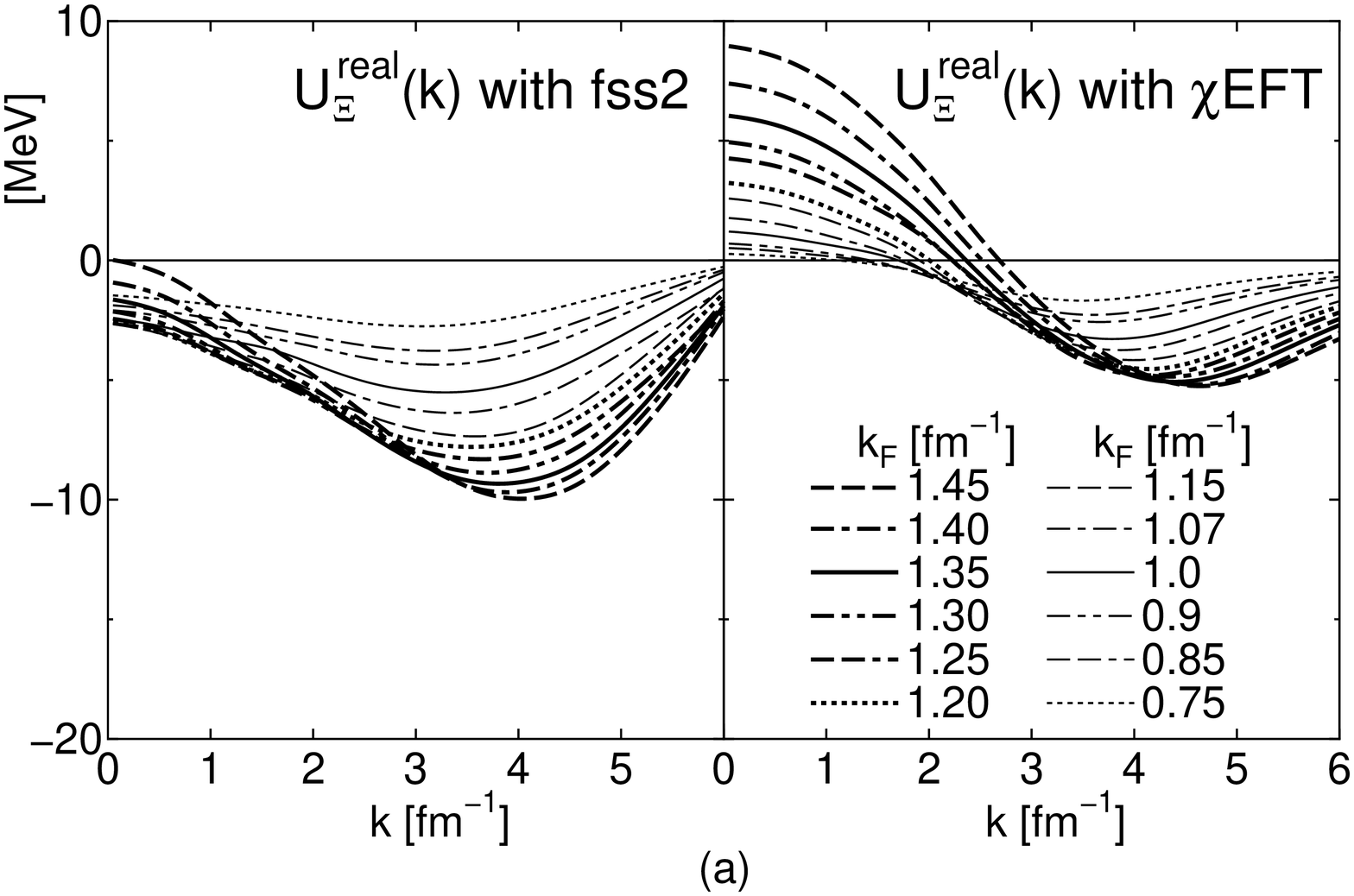} 
\caption{$\Lambda$ and $\Xi$ nuclear matter potentials calculated in the 
quark model \cite{fujiwara07} fss2 and in $\chi$EFT \cite{polinder06}. 
Figure adapted from Ref.~\citen{kohno10}.} 
\label{fig:lamxipots} 
\end{figure} 
\begin{figure} 
\centering 
\includegraphics[width=6.5cm]{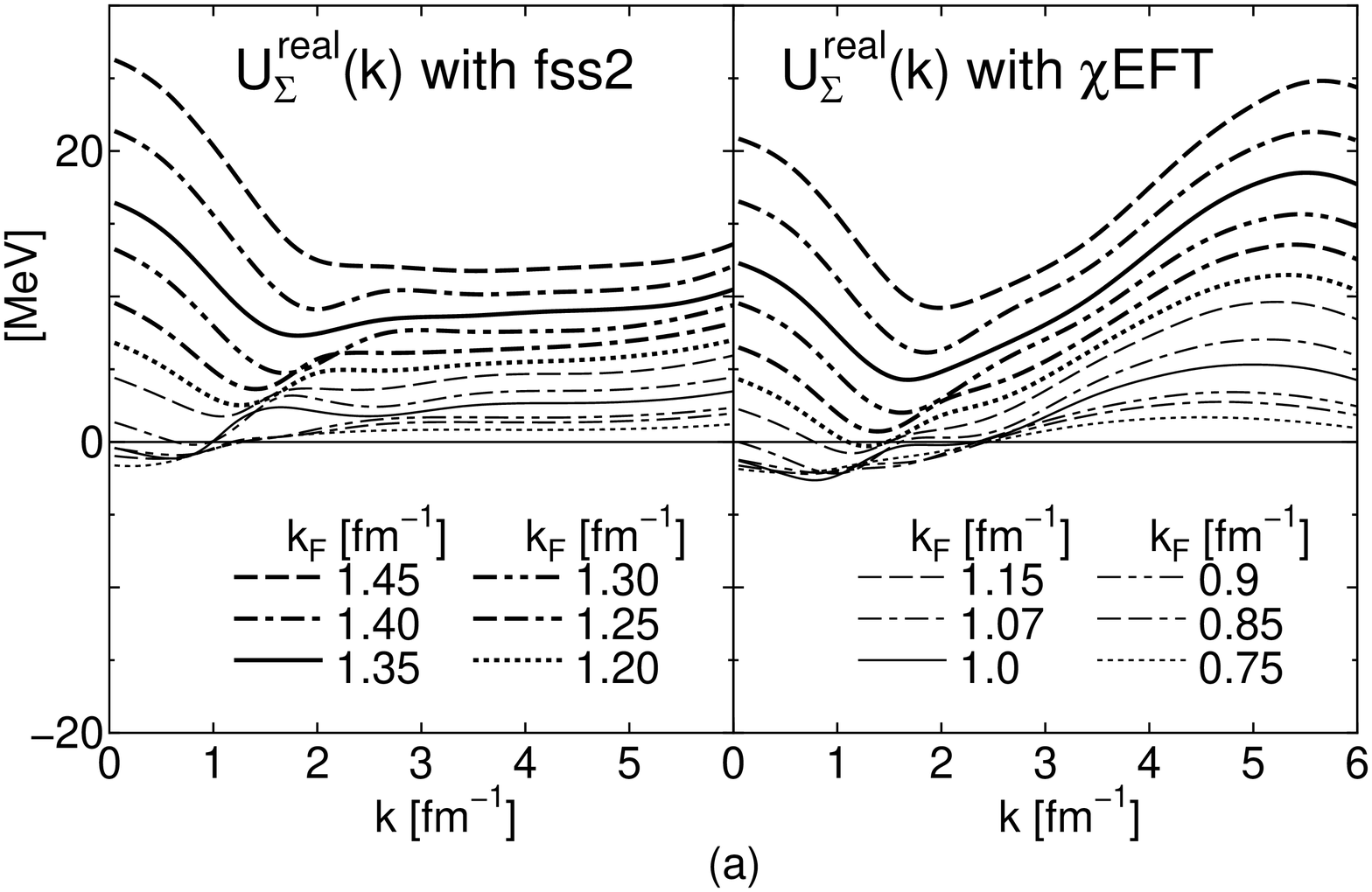} 
\hspace*{3mm} 
\includegraphics[width=6.5cm]{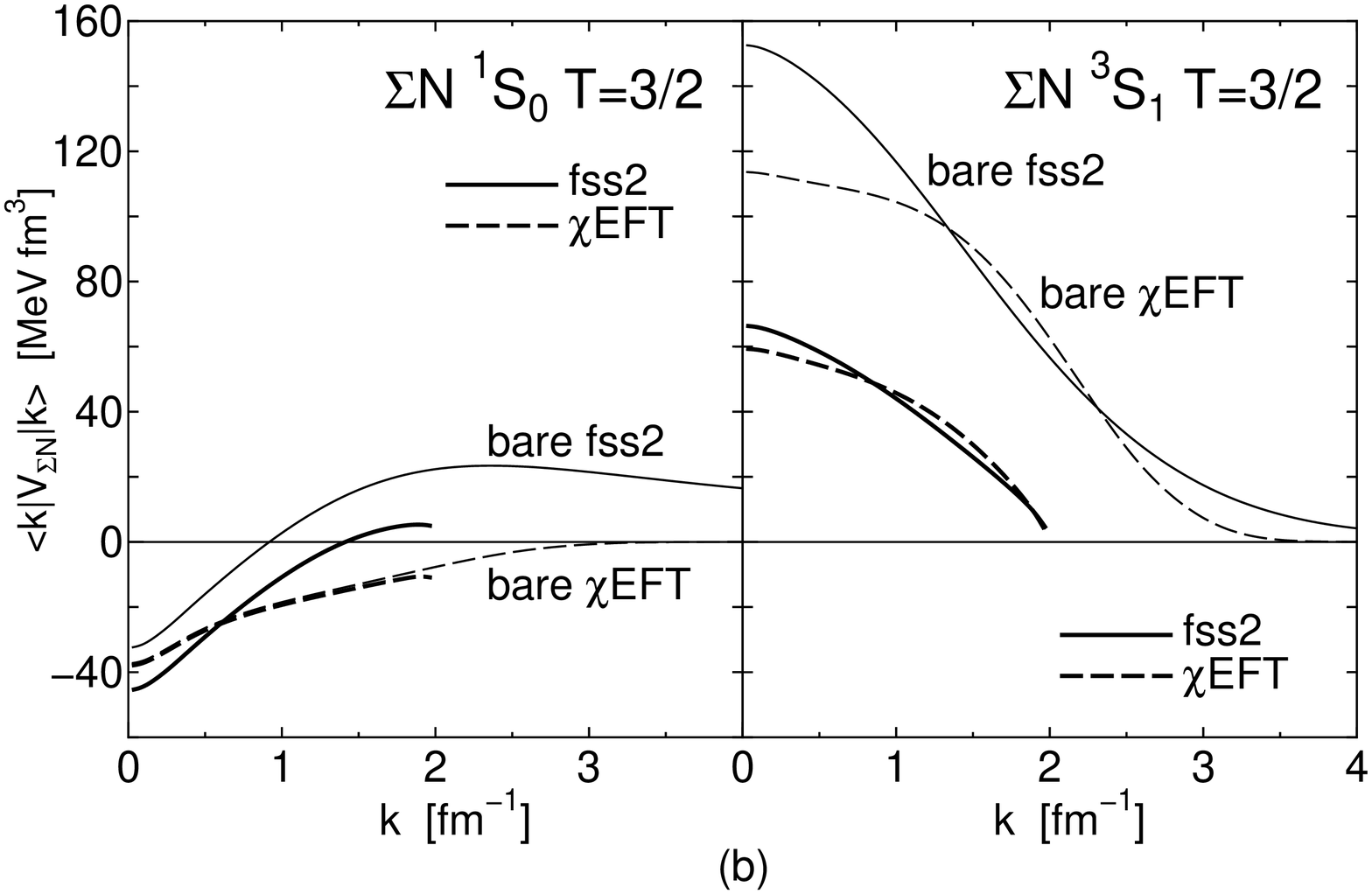} 
\caption{Left: $\Sigma$ nuclear matter potentials calculated in the 
quark model \cite{fujiwara07} fss2 and in $\chi$EFT \cite{polinder06}. 
Right: $T=3/2$ $\Sigma N$ potentials in these same models. Figure adapted 
from Ref.~\citen{kohno10}.}
\label{fig:sigpots} 
\end{figure} 

Very little is established experimentally on the interaction of $\Xi$ hyperons 
with nuclei. Inclusive $(K^-,K^+)$ spectra \cite{fukuda98} on $^{12}$C yield 
a somewhat shallow attractive potential, $V^{\Xi} \approx -14$~MeV, by fitting 
near the $\Xi^-$ hypernuclear threshold. All of the Nijmegen soft-core 
potentials listed in Table~\ref{tab:sig} produce a somewhat stronger 
isoscalar attraction, $V_0^{\Xi}\approx -25\pm 7$~MeV, while giving rise 
selectively, owing to the strong spin and isospin dependence which is 
reflected in the large size of the isovector $V_1^{\Xi}$ potential, to 
predictions of quasibound $\Xi$ states in several light nuclear targets, 
beginning with $^7$Li \cite{hiyama08}. These predictions should be considered 
with a grain of salt since no phenomenological constraint exists on 
$V_1^{\Xi}$. A `day-1' experiment at J-PARC on a $^{12}$C target has been 
scheduled \cite{tanida10}. 

It is worth noting that the main features provided by the Nijmegen potentials 
for hyperon-nuclear potentials also arise, at least qualitatively, in other 
models. This is demonstrated in Fig.~\ref{fig:lamxipots} for $\Lambda$ and 
$\Xi$ nuclear potentials, and in Fig.~\ref{fig:sigpots} for $\Sigma$ nuclear 
potentials. Whereas the $\Lambda$ nuclear potential is essentially attractive, 
and as deep as $\approx -30$~MeV, the $\Xi$ nuclear potential is weaker and 
could even turn repulsive. The $\Sigma$ nuclear potential is repulsive, as 
argued above. The r.h.s. of Fig.~\ref{fig:sigpots} demonstrates the origin of 
this repulsion, for $T=3/2$, due to the $\Sigma N$ $^3S_1$ channel which is 
strongly dominated by the Pauli exclusion principle for quarks at short 
distances. 

$\Xi$ hyperons could become stabilized in multi-$\Lambda$ hypernuclei 
once the decay $\Xi N \to \Lambda\Lambda$, which releases $\approx 25$~MeV 
in free space, gets Pauli blocked.{\footnote{With $\approx 80$~MeV release 
in $\Sigma N \to \Lambda N$, however, $\Sigma$ hyperons are unlikely to 
stabilize.}} The onset of $\Xi$ particle-stability would occur 
for $_{\Xi^0\Lambda}^{~~~6}$He or for $_{\Xi^0\Lambda\Lambda}^{~~~~7}$He, 
depending on whether or not $_{\Xi^0}^{~5}{\rm He}$ is bound, and by how much 
(if bound) \cite{sbg94}. It was shown that a large strangeness fraction, 
$-S/A\approx 0.7$ could be reached upon adding $\Xi$s to multi-$\Lambda$ 
hypernuclei in particle-stable configurations \cite{sbg93}, which leads to the 
concept of Strange Hadronic Matter (SHM) 
consisting of equal fractions of protons, neutrons, $\Lambda$, $\Xi^0$ and 
$\Xi^-$ hyperons \cite{sbg93}, with $f_S=1$ as in Strange Quark Matter 
(SQM). Both SHM and SQM provide macroscopic realizations of strangeness, 
but SHM is more plausible phenomenologically, whereas SQM is devoid of any 
experimental datum from which to extrapolate.

\section{${\overline K}$ nuclear interactions and ${\overline K}$ condensation} 
\label{sec:Kbar} 

The $\bar K$-nucleus interaction near threshold comes strongly attractive and 
absorptive in fits to the strong-interaction shifts and widths of $K^-$-atom 
levels~\cite{fg07}, resulting in deep potentials, 
Re~$V^{\bar K}(\rho_0)\sim -(150-200)$ MeV at threshold \cite{fgb93}. 
Chirally based coupled-channel models that fit the low-energy $K^-p$ 
reaction data, and the $\pi\Sigma$ spectral shape of the $\Lambda(1405)$ 
resonance, yield weaker but still very attractive potentials, 
Re~$V^{\bar K}(\rho_0)\sim -100$ MeV, as summarized recently in 
Ref.~\citen{weise08}. A third class, of relatively shallow potentials 
with Re~$V^{\bar K}(\rho_0) \sim -(40-60)$ MeV, was obtained by imposing 
a Watson-like self-consistency requirement \cite{ramoset00}. 

The onset of nuclear (quasi) binding for $K^-$ mesons occurs already with just 
one proton: the $\Lambda(1405)$ which is often represented by an $S$-matrix 
pole about 27 MeV below the $K^-p$ threshold. However, in chirally based 
models, the $I=0$ $\bar K N - \pi\Sigma$ coupled channel system exhibits also 
another $S$-matrix pole roughly 12 MeV below threshold and it is this pole 
that enters the effective $\bar KN$ interaction, affecting dominantly the 
$\bar K$-nucleus dynamics \cite{weise08}. The distinction between models that 
consider the twin-pole situation and those that are limited to the 
$\Lambda(1405)$ single-pole framework shows up already in calculations of 
$[\bar K (NN)_{I=1}]_{I=1/2,J^{\pi}=0^-}$, loosely denoted $K^-pp$, which 
is the configuration that maximizes the strongly attractive $I=0~\bar K N$ 
interaction with two nucleons. In Table~\ref{tab:kpp} which summarizes 
$K^-pp$ binding-energy calculations, the $I=0$ $\bar K N$ binding input to 
the first variational calculation is stronger by about 15 MeV than for the 
second one, resulting in almost 30 MeV difference in $B_{K^-pp}$. Furthermore, 
it is apparent from the `coupled-channel' entries in the table that the 
explicit use of the $\pi\Sigma N$ channel adds about $20 \pm 5$~MeV to the 
binding energy calculated using effective $\bar K N$ potential within 
a single-channel calculation. It is fair to state that in spite of the wide 
range of binding energies predicted for $K^-pp$, its existence looks robust 
theoretically. The experimental state of the art in searching for a $K^-pp$ 
signal is rather confused, as discussed during HYP09 [{\em Nucl. Phys. A} 
{\bf 835} (2010)]. New experiments, at GSI with a proton beam and at J-PARC 
with pion and with kaon beams are underway \cite{nagae10}. 

\begin{table} 
\caption{Calculated $B_{K^-pp}$, mesonic ($\Gamma_{\rm m}$) 
\& nonmesonic ($\Gamma_{\rm nm}$) widths.}
\label{tab:kpp} 
\begin{center} 
\begin{tabular}{lccccc} \hline\hline 
&\multicolumn{2}{c}{$\bar K NN$ single channel} 
&\multicolumn{3}{c}{$\bar K NN - \pi\Sigma N$ coupled channels} \\ 
(MeV) & variational~\cite{akaishi02} & variational~\cite{dote08} & 
Faddeev~\cite{shevch07} & Faddeev~\cite{ikeda07} & variational~\cite{wycech09} 
\\ \hline 
$B_{K^-pp}$ & $48$ & $17-23$ & $50-70$ & $60-95$ & $40-80$ \\ 
$\Gamma_{\rm m}$ & $61$ & $40-70$  & $90-110$ & $45-80$ & $40-85$ \\ 
$\Gamma_{\rm nm}$ & $12$ & $4-12$ & & & $\sim 20$ \\ \hline 
\end{tabular} 
\end{center} 
\end{table} 

\begin{figure}[tbh] 
\centering
\includegraphics[width=6.5cm]{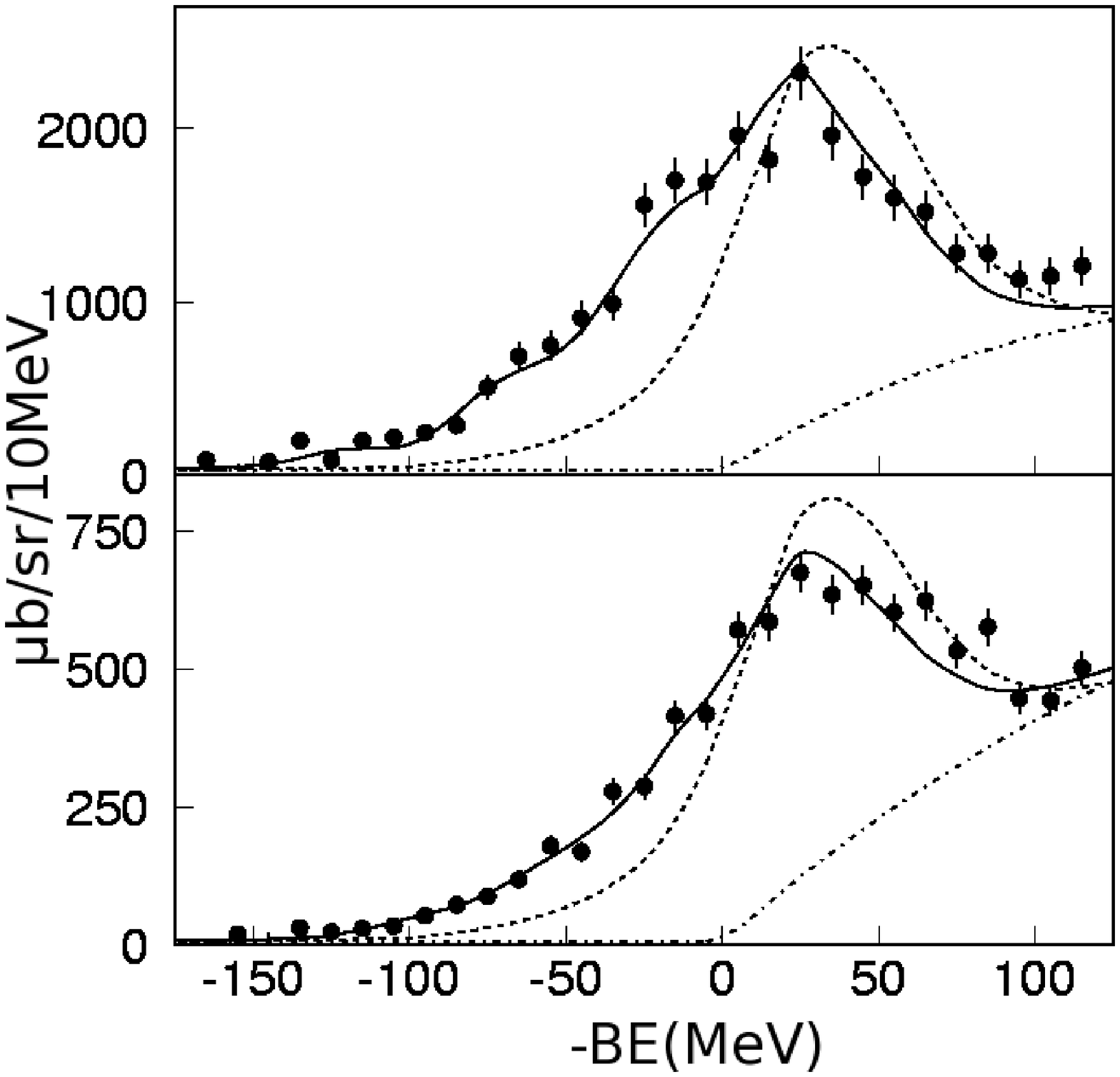} 
\hspace*{3mm} 
\includegraphics[width=6.5cm]{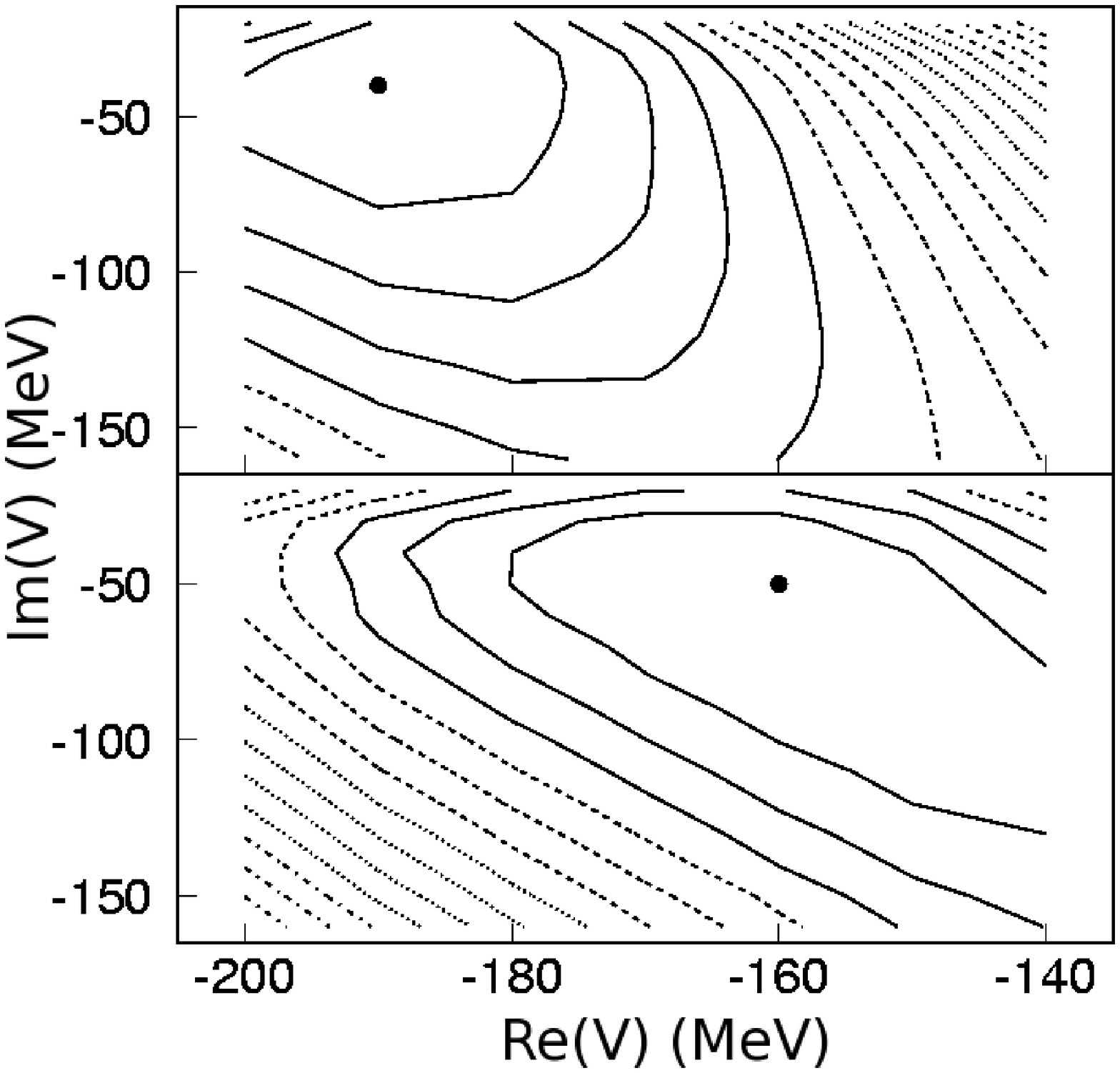} 
\caption{Missing mass spectra (left) and $\chi^2$ contour plots (right) for 
the inclusive reactions $(K^-,n)$ (upper) and $(K^-,p)$ (lower) at 
$p_{K^-}=1$~GeV/c on $^{12}$C, from Ref.~\citen{kish07}} 
\label{fig:kish} 
\end{figure} 

A fairly new and independent evidence in favor of deep 
$\bar K$-nucleus potentials is provided by $(K^-,n)$ and $(K^-,p)$ 
spectra \cite{kish09} taken at KEK on $^{12}$C, and very recently also 
on $^{16}$O at $p_{K^-}=1$ GeV/c. The $^{12}$C 
spectra are shown in Fig.~\ref{fig:kish}, where the solid lines on the 
left-hand side represent calculations (outlined in Ref.~\citen{yamagata06}) 
using potential depths in the range $160-190$ MeV. The dashed lines 
correspond to using relatively shallow potentials of depth about 60 MeV 
which I consider therefore excluded by these data.{\footnote{This conclusion, 
for the $(K^-,p)$ spectrum, has been disputed recently by Magas {\it et al.} 
\cite{magas10}}} Although the potentials that fit these data are sufficiently 
deep to support strongly-bound antikaon states, a fairly sizable extrapolation 
is required to argue for $\bar K$-nuclear quasibound states at energies of 
order 100 MeV below threshold, using a potential determined largely near 
threshold. Furthermore, the best-fit Im~$V^{\bar K}$ depths of $40-50$ MeV 
imply that $\bar K$-nuclear quasibound states are broad, as studied in 
Refs.~\citen{mares06,gazda07}. 

\begin{figure}[tbh] 
\centering
\includegraphics[width=6.5cm]{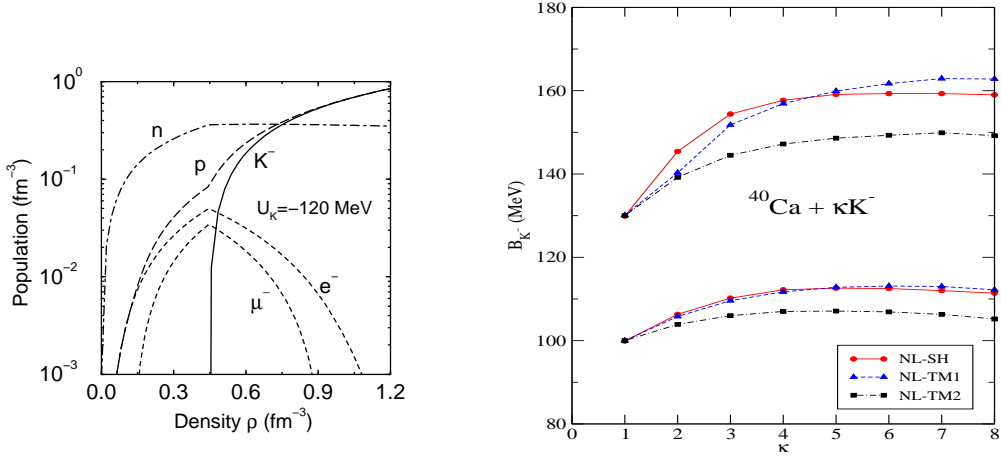} 
\hspace*{3mm} 
\includegraphics[width=6.5cm,height=6.0cm]{multik40ca.eps} 
\caption{Left: calculated neutron-star population as a function of density, 
from Ref.~\citen{glend99}. The neutron density stays nearly constant once 
kaons condense. Right: calculated separation energies $B_{K^-}$ in multi-$K^-$ 
nuclei based on $^{40}{\rm Ca}$ as a function of the number $\kappa$ of 
$K^-$ mesons in several RMF models, for two choices of parameters fixed 
for $\kappa=1$, from Ref.~\citen{gazda08}.}
\label{fig:gfgm} 
\end{figure} 

A robust consequence of the sizable $\bar K$-nucleus attraction is that 
$K^-$ condensation, when hyperon degrees of freedom are ignored, could 
occur in neutron star matter at about 3 times nuclear matter density, 
as shown on the l.h.s. of Fig.~\ref{fig:gfgm}. Comparing it with the 
r.h.s. of Fig.~\ref{fig:sbg}, for neutron stars, but where strangeness 
materialized through hyperons, one may ask whether $\bar K$ mesons condense 
also in SHM. This question was posed and answered negatively long time ago 
for neutron star matter, but only recently for SHM in Ref.~\citen{gazda08} 
by calculating multi-$\bar K$ nuclear configurations. The r.h.s. of 
Fig.~\ref{fig:gfgm} demonstrates a remarkable saturation of $K^-$ separation 
energies $B_{K^-}$ calculated in multi-$K^-$ nuclei, independently of the 
applied RMF model. 
The saturation values of $B_{K^-}$ do not allow conversion of hyperons 
to $\bar K$ mesons through the strong decays $\Lambda \to p + K^-$ 
or $\Xi^- \to \Lambda + K^-$ in multi-strange hypernuclei, which therefore 
remain the lowest-energy configuration for multi-strange systems. This 
provides a powerful argument against $\bar K$ condensation in the laboratory, 
under strong-interaction equilibrium conditions \cite{gazda08}. It does not 
apply to kaon condensation in neutron stars, where equilibrium configurations 
are determined by weak-interaction conditions. This work has been recently 
generalized to multi-$K^-$ {\it hypernuclei} \cite{gazda09}.

\section*{Acknowledgments} 
Thanks are due to John Millener for instructive correspondence on the analysis 
of the $\gamma$-ray experiments, and to the Organizers of the NFQCD10 Workshop 
at the Yukawa Institute, Kyoto, for their generous hospitality.

\end{document}